\documentclass[preprint,showpacs,preprintnumbers,amsmath,amssymb]{revtex4}

\usepackage{graphicx}
\usepackage{dcolumn}
\usepackage{bm}

\begin{document}

\preprint{APS/123-QED}

\title{Effect of symmetry energy on nuclear stopping and its relation to the 
production of light charged fragments\\}

\author{Sanjeev Kumar}
\author{Suneel Kumar}%
 \email{suneel.kumar@thapar.edu}
\affiliation{%
School of Physics and Material Science, Thapar University, Patiala-147004, Punjab (India)\\
}%

\author{Rajeev K. Puri}
\affiliation{
Department of Physics, Panjab University, Chandigarh-160014 (India)
}%

\date{\today}

\begin{abstract}
We present a complete systematics (excitation function, impact parameter, system size, isospin asymmetry
and equations of state dependences) of global stopping and fragments production for heavy-ion reactions 
in the energy range between 50 and 1000 MeV/nucleon in the presence of symmetry energy
and isospin-dependent cross-section.  It is observed that the degree of stopping depends weakly on the symmetry
energy and strongly on the isospin-dependent cross-section. On the other hand, the symmetry energy and
isospin-dependent cross-section has an effect 
of the order of more than 10$\%$ on the emission of light charged particles (LCP's). It means that nuclear 
stopping and LCP's can be used as a tool to get the information of isospin-dependent cross-section. 
Interestingly, the LCP's emission in the presence of symmetry energy is found to be highly correlated with the global
stopping.\\
\end{abstract}

\maketitle

\section{Introduction}

One of the goals of heavy-ion collisions (HIC) at intermediate energies is to extend the knowledge 
of hot and dense nuclear
matter to the extreme conditions. In the past, these studies were focused on multifragmentation, 
that constitutes fragments of all sizes\cite{Sing00}. 
Additional promising observable  for the understanding of nuclear equation of state is 
the anisotropy in the 
momentum distribution
that includes the directed in-plane flow (bounce off ) as well as out of plane flow (squeeze out) \cite{Stoc86,Aich91}. The absolute values of the flow results from the 
interplay between the attractive mean field and repulsive nucleon-nucleon scatterings.
This interplay is also responsible for the transition from a fused state to total disassemble 
one. 
The another phenomena
linked with the above interplay is the global stopping of nuclear matter. 
Recently, Puri and co-workers\cite{Dhaw06}, 
tried to correlate the multifragmentation with global nuclear stopping. 
Their findings revealed that light charged particles acts in a similar fashion 
like anisotropy ratio. They, however, did not take isospin of the system into account.\\ 
Following the recent development of radioactive beam facilities in many parts of the world, it 
became possible to study the neutron (or proton) rich nuclear collisions at intermediate energies. Therefore, for a meaningful investigation, one should include the isospin 
dependence of the field. As pointed out by Bauer \cite{Baur88}, nuclear stopping at intermediate energies is 
determined by the mean field as well as by the in-medium NN cross-sections. 
Unfortunately, his calculations were silent about the symmetry potential. 
The recent work of many authors \cite{Li95,Liu01,Feng02} suggest that the 
degree of approaching isospin equilibration helps 
to probe the nuclear stopping in heavy-ion collisions. 
In Ref.\cite{Liu01}, isospin dependence of cross-section
was investigated in nuclear stopping. In a recent communication \cite{Feng02}, authors studied
the behavior of excitation function $Q_{zz}/nucleon$ and concluded that 
$Q_{zz}/nucleon$ can provide 
information about the isospin dependence in term of cross-sections. 
Several more studies have also focused in the recent years on the 
isospin degree of freedom \cite{Liu04}.\\
We wish to focus on the systematic study of isospin dependence and will focus 
on the relation between light charged particles and equilibration of the reaction using
isospin-dependent quantum molecular dynamics. \\
This study is done within the framework of isospin-dependent quantum molecular dynamics model
that is explained in section-II. The results are presented in section-III. We present summary in section-IV.\\

\section{ISOSPIN-DEPENDENT QUANTUM MOLECULAR DYNAMICS (IQMD) MODEL}
The isospin-dependent quantum molecular dynamics (IQMD)\cite{Hart89} model treats different charge states of
nucleons, deltas and pions
explicitly\cite{Hart03}, as inherited from the VUU model \cite{Krus85}. The IQMD model has been used successfully
for the analysis of large number of observables from low to relativistic energies 
\cite{Hart89,Hart03,Hart02}.
The isospin degree of
freedom enters into the calculations via both cross-sections and 
mean field\cite{Krus85,Bass95}.
The details about the elastic and inelastic cross-sections
for proton-proton and neutron-neutron collisions can be found in Ref.\cite{Hart89,Hart02}. \\
In this model,  baryons are represented by Gaussian-shaped density distributions
\begin{equation}
f_i(\vec{r},\vec{p},t) = \frac{1}{\pi^2\hbar^2}~e^{-(\vec{r}-\vec{r_i}(t))^{2}\frac{1}{2L}}~e^{-(\vec{p}-\vec{p_i}(t))^{2}\frac{2L}{\hbar^2}}.
\end{equation}
Nucleons are initialized in a sphere with radius $R= 1.12 A^{1/3}$ fm, in accordance with the liquid drop model. Each nucleon occupies a volume of $h^3$, so that phase space is uniformly filled. The initial momenta are randomly chosen between 0
and Fermi momentum($p_F$). The nucleons of target and projectile
interact via two and three-body Skyrme forces and Yukawa potential. The
isospin degree of freedom is treated explicitly by employing a symmetry potential and explicit Coulomb forces
between protons of colliding target and projectile. This helps in achieving correct distribution of protons and neutrons
within nucleus.\\
The hadrons propagate using Hamilton equations of motion:
\begin{equation}
\frac{d{r_i}}{dt}~=~\frac{d\it{\langle~H~\rangle}}{d{p_i}}~~;~~\frac{d{p_i}}{dt}~=~-\frac{d\it{\langle~H~\rangle}}{d{r_i}},
\end{equation}
with
\begin{eqnarray}
\langle~H~\rangle&=&\langle~T~\rangle+\langle~V~\rangle\nonumber\\
&=&\sum_{i}\frac{p_i^2}{2m_i}+
\sum_i \sum_{j > i}\int f_{i}(\vec{r},\vec{p},t)V^{\it ij}({\vec{r}^\prime,\vec{r}})\nonumber\\
& &\times f_j(\vec{r}^\prime,\vec{p}^\prime,t)d\vec{r}d\vec{r}^\prime d\vec{p}d\vec{p}^\prime .
\end{eqnarray}
 The baryon-baryon potential $V^{ij}$, in the above relation, reads as:
\begin{eqnarray}
V^{ij}(\vec{r}^\prime -\vec{r})&=&V^{ij}_{Skyrme}+V^{ij}_{Yukawa}+V^{ij}_{Coul}+V^{ij}_{sym}\nonumber\\
&=& \left [t_{1}\delta(\vec{r}^\prime -\vec{r})+t_{2}\delta(\vec{r}^\prime -\vec{r})\rho^{\gamma-1}
\left(\frac{\vec{r}^\prime +\vec{r}}{2}\right) \right]\nonumber\\
& & +~t_{3}\frac{exp(|\vec{r}^\prime-\vec{r}|/\mu)}{(|\vec{r}^\prime-\vec{r}|/\mu)}~+~\frac{Z_{i}Z_{j}e^{2}}{|\vec{r}^\prime -\vec{r}|}\nonumber\\
& & + t_{6}\frac{1}{\varrho_0}T_3^{i}T_3^{j}\delta(\vec{r_i}^\prime -\vec{r_j}).
\label{s1}
\end{eqnarray}
Here $Z_i$ and $Z_j$ denote the charges of $i^{th}$ and $j^{th}$ baryon, and $T_3^i$, $T_3^j$ are their respective $T_3$
components (i.e. 1/2 for protons and -1/2 for neutrons). Meson potential consists of Coulomb interaction only.
The parameters $\mu$ and $t_1,.....,t_6$ are adjusted to the real part of the nucleonic optical potential. For the density
dependence of nucleon optical potential, standard Skyrme-type parameterization is employed.
The choice of equation of state (or compressibility) is still controversial one. Many studies
advocate softer matter, whereas, much more believe the matter to be harder in nature 
\cite{Krus85,Mage00}. 
We shall use both hard (H) and soft (S) equations of state that have 
compressibilities of 380 and 200 MeV, 
respectively.\\

The binary nucleon-nucleon collisions are included by employing the collision 
term of well known VUU-BUU equation
\cite{Stoc86,Krus85}. The binary collisions
are done stochastically, in a similar way as are done in all transport models. During the propagation, two nucleons are
supposed to suffer a binary collision if the distance between their centroids
\begin{equation}
|r_i-r_j| \le \sqrt{\frac{\sigma_{tot}}{\pi}}, \sigma_{tot} = \sigma(\sqrt{s}, type),
\end{equation}
"type" denotes the ingoing collision partners (N-N, N-$\Delta$, N-$\pi$,..). In addition,
Pauli blocking (of the final
state) of baryons is taken into account by checking the phase space densities in the final states.
The final phase space fractions $P_1$ and $P_2$ which are already occupied by other nucleons are determined for each
of the scattering baryons. The collision is then blocked with probability
\begin{equation}
P_{block}~=~1-(1-P_1)(1-P_2).
\end{equation}
The delta decays are checked in an analogous fashion with respect to the phase space of the resulting nucleons.\\

\section{Results and Discussion}
\begin{figure}
\includegraphics{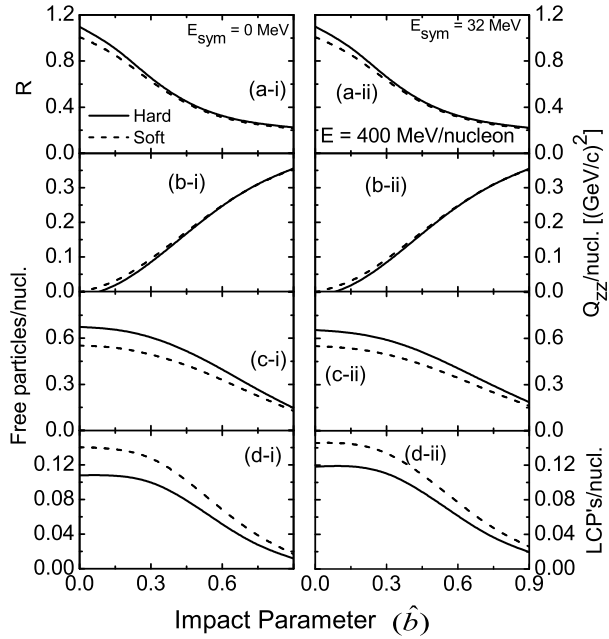}
\caption{\label{fig:5} Impact parameter dependence of (a) R, (b) ${Q_{zz}/nucleon}$, (c) free particles/nucleon,
 (d) LCP's/nucleon, with hard (H) and soft (S) equations of state. The results are displayed at $E_{Sym} = 0$ (i)
and 32 MeV (ii).}
\end{figure}

The global stopping in heavy-ion collisions has been studied with the help of many different variables. In earlier studies, one used to relate the rapidity distribution with 
global stopping. The rapidity distribution can be defined as
\cite{Dhaw06,Wong94}:
\begin{equation}
Y(i)= \frac{1}{2}ln\frac{E(i)+p_{z}(i)}{E(i)-p_{z}(i)},
\end{equation}
where $E(i)$ and $p_z(i)$ are, respectively, the total energy and longitudinal momentum of $i^{th}$ particle. For a
complete stopping, one expects a single Gaussian shape. Obviously, narrow Gaussian indicate 
better thermalization
compared to broader Gaussian.\\
The second possibility to probe the degree of stopping is the anisotropy ratio (R) \cite{Liu01}:
\begin{equation}
R = \frac{2}{\pi}\frac{\left(\sum_{i}|p_{\perp}(i)|\right)}{\left(\sum_{i}|p_{\parallel}(i)|\right)},
\end{equation}
where, summation runs over all nucleons. The transverse and longitudinal momenta 
are $p_{\perp}(i)$ =
$\sqrt{p_x^2(i) + p_y^2(i)}$ and $p_{\parallel}(i) = p_z(i)$, respectively.
Naturally, for a complete stopping,
$R$  should be close to unity.\\
Another quantity, which is indicator of nuclear stopping and has been used recently, 
is the quadrupole moment $Q_{zz}$, defined as\cite{Liu01}:
\begin{equation}
Q_{zz} = \sum_{i} \left(2p_z^2(i) - p_x^2(i) - p_{y}^2(i)\right).
\end{equation}
Naturally, for a complete stopping, $Q_{zz}$ should be close to 0.\\
In the present analysis, thousands of event were simulated for the neutron-rich reaction of 
$_{54}Xe^{131}~+~_{54}Xe^{131}$
at incident energies between 50 and 1000 MeV/nucleon using hard 
equation of state along with energy dependent Cugnon and constant
nucleon-nucleon cross-sections\cite{Kuma98}. Moreover, to see the effect of compressibilities on nuclear 
stopping and fragmentation, soft equation of state is also used in Fig.\ref{fig:5}. 
The geometry of the collision was varied between
the most-central to peripheral one. The role of symmetry energy is studied by simulating the above reaction
with and without this term. As stated above, we plan to study the degree of stopping and 
emission of 
fragments using symmetry energy and isospin-dependent cross-section. We shall also 
correlate the degree of stopping with the emission of light charged particles
as is also done in Ref.\cite{Dhaw06}. 
The fragments are constructed within
minimum spanning tree (MST) method \cite{Sing00}, which binds nucleons if they are with in a distance 
of 4 fm.\\
\begin{figure}
\includegraphics{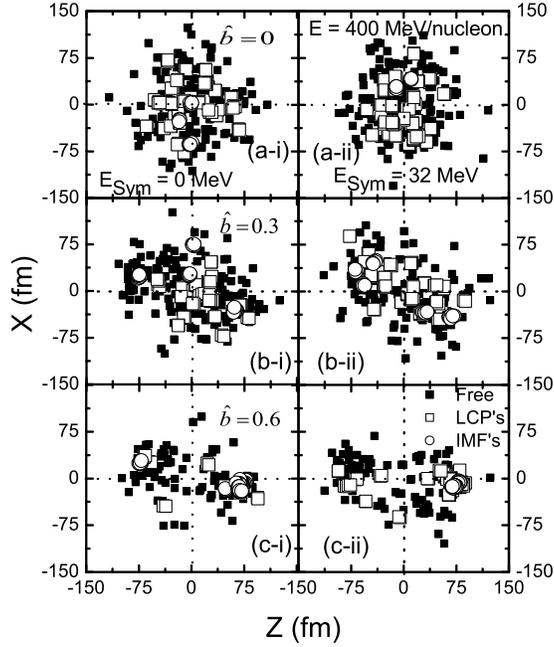}
\caption{\label{fig:1}  The final phase space of a single event for the reaction of
$_{54}Xe^{131}~+~_{54}Xe^{131}$ with(ii)
and without symmetry energy(i). The top(a), middle(b) and bottom (c) panels are, respectively, for scaled
impact parameters $\hat{b}$
= 0, 0.3, 0.6. Different symbols are for free nucleons, LCP's and IMF's.}
\end{figure}
In Fig.\ref{fig:1}, 
we display the final phase space of a single event of $_{54}Xe^{131}~+~_{54}Xe^{131}$ at incident
energy of 400
MeV/nucleon, with and without symmetry energy. The top, middle and bottom panels are at 
$\hat{b}$ = 0, 0.3 and 0.6, respectively. Here phase space of free particles [A =1], light charged particles (LCP's) 
[2$\le$A$\le$4] and intermediate mass fragments(IMF's)[5$\le$ A $\le$ 44] is displayed. We note that 
irrespective of the symmetry energy, central collisions
lead to complete spherical distribution of particles, indicating, spreading of 
the nucleons in all directions. It means that breaking of initial correlations among nucleons is maximal in this
region and, as a result, more
randomization and stopping in the hot and compressed nuclear matter occurs. 
This effect seems to decrease with impact parameter. 
Since free particles as well as LCP's originate from the 
mid-rapidity region, they are better suited for studying the degree of stopping 
reached in a heavy-ion collision. 
On the other hand, IMF's seems to originate
either from the target or from the projectile region, therefore, are 
the remnant/residue of the spectator matter. This observation is 
in agreement with many other studies \cite{Sing00,Dhaw06,Dhaw07}.\\
\begin{figure}
\includegraphics{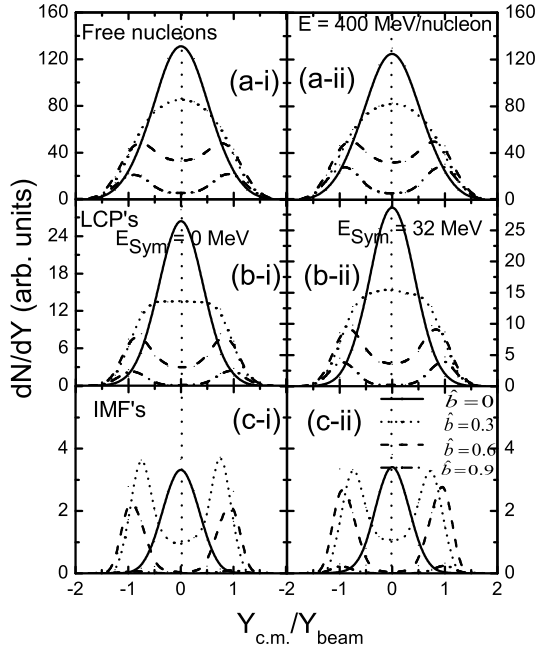}
\caption{\label{fig:2} The rapidity distribution $\frac{dN}{dY}$ as a function of
reduced rapidity for free nucleons(a), LCP's (b) and IMF's(c)
at different impact parameters. The reaction under study is $_{54}Xe^{131}~+~_{54}Xe^{131}$ at incident energy E = 400
MeV/nucleon. The left and right panels are with (ii) and without symmetry energy (i).}
\end{figure}
\begin{figure}
\includegraphics{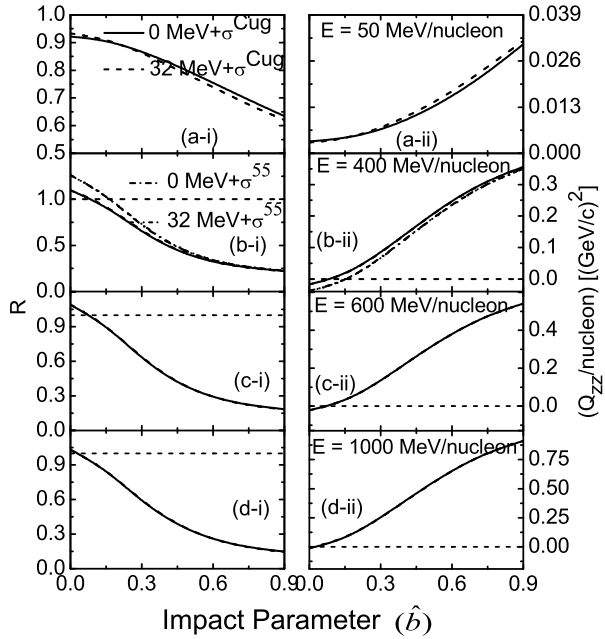}
\caption{\label{fig:3} The anisotropy ratio $R$ (i) and quadrupole moment
$\left({Q_{zz}/nucleon}\right)$ (ii) as a function
of normalized impact parameter with and without symmetry energy. In panel (b), the results are also
displayed with isospin-dependent cross-section (55mb). The panels from top to bottom are at incident energy
of 50 (a), 400 (b), 600 (c) and 1000 MeV/nucleon (d), respectively.}
\end{figure}

\begin{figure}
\includegraphics{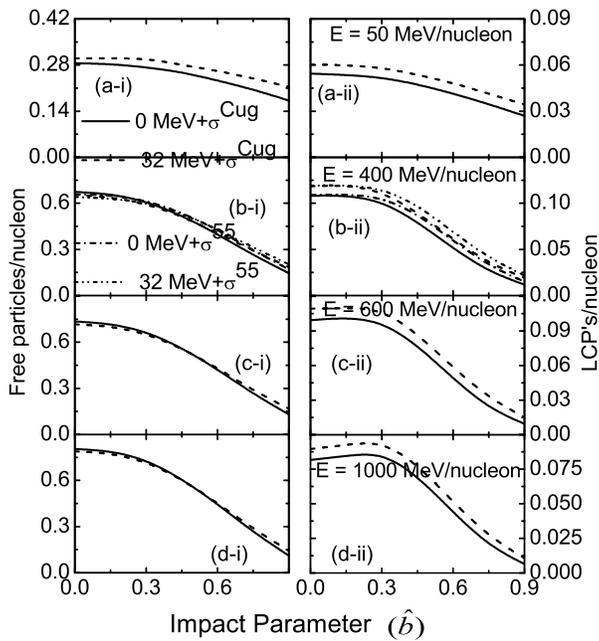}
\caption{\label{fig:4} Same as in Fig.\ref{fig:3},
but for the multiplicity of free particles/nucleon and LCP's/nucleon.}
\end{figure}

To further quantify this observation, we display in Fig.\ref{fig:2}, 
the rapidity distribution $\frac{dN}{dY}$ for 
the emission of free nucleons as well as LCP's and IMF's. 
We see that free particles and LCP's emitted in the central 
collisions form a single narrow Gaussian shape, whereas, IMF's have broader 
Gaussian, indicating less thermalization. As we increase impact parameter, single
Gaussian distribution splits into two Gaussian (at target and projectile rapidities), 
indicating correlated matter. From the shape of the Gaussian, 
one sees that free particles {and LCP's} are better indicator of thermal source. 
Obviously, this condition is necessary, 
but, not a sufficient one.\\
From the figure, it is also evident that the symmetry energy does not plays 
significant role for the rapidity 
distribution. The peak value of the Gaussian for LCP's is altered
by about 10$\%$, whereas, nearly no effect is seen in the case of intermediate mass fragments.
The reason is that LCP's can feel the role of mean field directly, while, the heavy fragments have weak sensitivity
\cite{Yan07}. 
From the figure, one sees a one to one relation between the degree of stopping and emission 
of LCP's. These conclusions match with the findings of Fig.\ref{fig:1} 
and 
Ref.\cite{Dhaw06}.\\
In Fig.\ref{fig:3}, 
we display impact parameter dependence of global
variables ($R$ and 
$Q_{zz}/nucleon$), whereas,  the multiplicity dependence of free nucleons
and LCP's is displayed in Fig.\ref{fig:4}.  
The displayed results are at $E_{Sym}$= 0 
and $E_{Sym}$ = 32 MeV in each panel, while, in panel (b) the results are also displayed
with isospin-dependent cross-section. The value of cross-section is denoted in the superscript.
From Fig.\ref{fig:3}, 
We observe that $R$ and ${Q_{zz}/nucleon}$ behave in opposite fashion i.e. $R$ and $\frac{1}{Q_{zz}/nucleon}$ will 
behave in a similar fashion.
For $R > 1$ and $Q_{zz}/nucleon < 0$, it can
be explained by the preponderance of momentum flow perpendicular to the beam 
direction\cite{Renf84}. The maximum stopping is observed around 400 MeV/nucleon, which is in 
supportive nature with the findings of W. Reisdorf {\it et al.}\cite{Reis04a}.
In their work, they measured the nuclear stopping from 0.090 to 1.93 GeV/nucleon and maximal stopping
was observed around 400 MeV/nucleon. It is clear that if the reaction reaches the maximal stopping 
around certain energies, the matter formed in the reaction should reach minimum transparency and thus
most of the particles are preferentially out-of-plane. On the other hand, no visible effect is seen for symmetry energy 
term. We see both quantities are nearly independent of the symmetry energy, while, strongly depends on
the isospin-dependent cross-section.\\
As we know, major contribution for the stopping of nuclear matter is from the hot 
and compressed region, where symmetry energy does not play any role. Some small
spikes can be seen at lower beam energies, however, outcome is independent of
the symmetry energy at higher incident energies. This is due to the fact that above the 
Fermi energy, incident energy itself is sufficient to break the 
initial correlations among the nucleons. On the other hand, isospin-dependent cross-section will lead to 
violent N-N collisions, which further cause the transformation of the initial longitudinal motion in 
other directions and hence thermalization of the system. This dominant role played by the isospin-dependent cross-section
gradually disappears with increase in the impact parameter. As discussed earlier, stopping is the phenomena which
originates from the participant zone and this zone goes on decreasing with increase in the impact parameter and hence the
effect of cross-section on nuclear stopping. 
These findings are also in supportive nature with the findings of Liu {\it et al.,} \cite{Liu01}. \\

To correlate the degree of stopping with the multiplicity of fragments, 
we display in Fig.\ref{fig:4}, 
the impact parameter
dependence of the multiplicity of free nucleons as well as of LCP's. 
The behavior of all curves is similar to that of nuclear stopping parameters $R$ and $\frac{1}{Q_{zz}/nucleon}$, as 
discussed in Fig.\ref{fig:3}. 
In addition, 
LCP's are more sensitive towards symmetry energy compared to free particles. Due to
pairing nature of LCP's, symmetry energy term $\propto$ $(N-Z)^2$ contributes considerably.
The effect of isospin-dependent cross-section is more visible for the LCP's as compared to free particles. 
This also gives us clue that LCP's production can act as a indicator for the nuclear stopping.  
Moreover, free particles/nucleon are found to 
increase monotonically with the incident energy, while LCP's/nucleon 
behave in similar fashion as that of nuclear stopping i.e. maximum around 400 MeV/nucleon and then decreases.  
It is also evident from Ref.\cite{Dhaw06}, LCP's production 
act as a barometer for nuclear stopping compared to the free particles.\\
\begin{figure}
\includegraphics{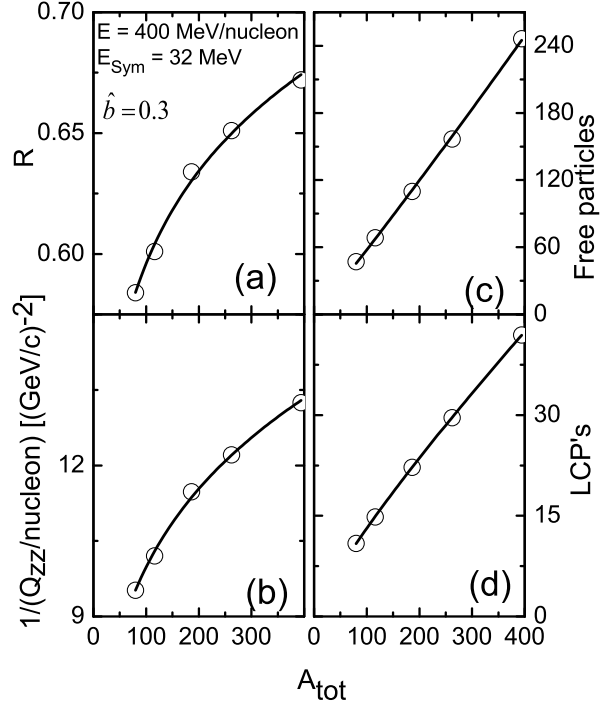}
\caption{\label{fig:6} System size dependence of (a) R, (b) $\frac{1}{Q_{zz}/nucleon}$, (c) free particles,
 (d) LCP's, in the presence of symmetry energy. All the curves are fitted with power law.}
\end{figure}

\begin{figure}
\includegraphics{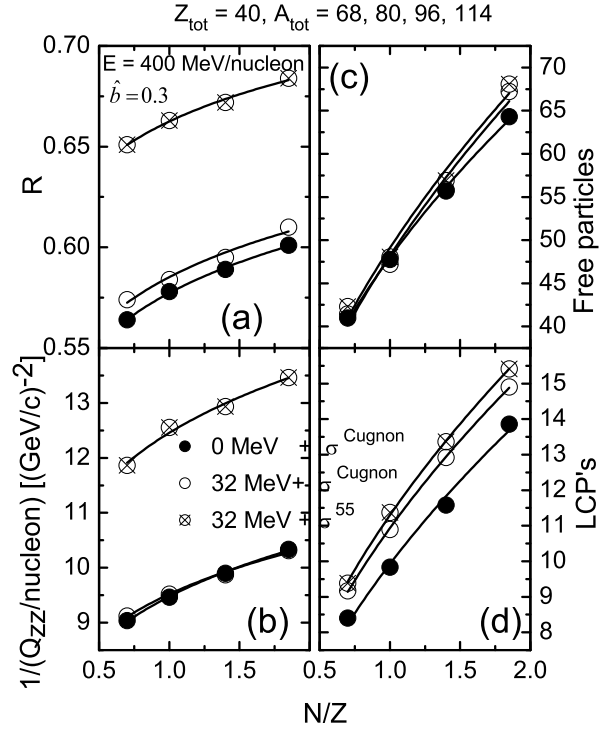}
\caption{\label{fig:7} Isospin asymmetry of  (a) R, (b) $\frac{1}{Q_{zz}/nucleon}$, (c) 
free particles, (d) LCP's, in the presence of symmetry energy and isospin-dependent 
cross-section (55 mb).}
\end{figure}

In Fig.\ref{fig:5}, 
we have checked the sensitivity of nuclear stopping as well as fragment production
with the nuclear equation of state (EOS). For this purpose, a hard (H) and soft (S) equations of state 
with compressibility $\kappa$ = 380, 200 MeV are employed, respectively. The nuclear stopping is found to be 
weakly dependent on the equations of state, while, the fragments production is sensitive to different equations of 
state. It means that the fragment production with different equations of state can act as a global indicator 
for the nuclear stopping as it is weakly dependent on equations of state.\\
It also becomes important to study the system size dependence and isospin asymmetry of R, $\frac{1}{Q_{zz}/nucleon}$,
free particles and LCP's. For this, in Fig.\ref{fig:6}, 
we have displayed the results for the reactions of 
$_{20}Ca^{40}~+~_{20}Ca^{40}$,  $_{28}Ni^{58}~+~_{28}Ni^{58}$, $_{41}Nb^{93}~+~_{41}Nb^{93}$,
 $_{54}Xe^{131}~+~_{54}Xe^{131}$ and  $_{79}Au^{197}~+~_{79}Au^{197}$, in which Z as well as A is varied.
On the other hand, results are displayed, in Fig.\ref{fig:7}, 
for the reactions of 
 $_{20}Ca^{34}~+~_{20}Ca^{34}$ (N/Z = 0.7),  $_{20}Ca^{40}~+~_{20}Ca^{40}$ (N/Z = 1),  
$_{20}Ca^{48}~+~_{20}Ca^{48}$ (N/Z = 1.4) and $_{20}Ca^{57}~+~_{20}Ca^{57}$ (N/Z = 1.85), having same Z and different A,
in the presence of symmetry energy and isospin-dependent cross-section. The curves in 
Figs.\ref{fig:6} 
and \ref{fig:7} 
 are parametrized with the power law $Y = CX^{\tau}$, where C and $\tau$ are constants, while X and Y
are the respective parameters on X and Y axis.\\

From the Fig.\ref{fig:6},  
it is observed that the parameters R, $\frac{1}{Q_{zz}/nucleon}$, free particles as 
well as LCP's are in similar trend with the composite mass of the system. All the parameters are found to increase
with the composite mass of the system. For a fixed geometry (semi-central here), more heavier is the composite
system, more hot is the compressed zone, which further results in more thermalization or 
global stopping. Looking the parallel side, the free particles and LCP's will always originate from the participant
zone. With an increase in the composite mass of the system, the participant zone goes on 
increasing
for a fixed geometry (semi-central here) and hence the production of free particles and LCP,s. Similar findings are also 
published in the Ref.\cite{Liu01, Reis04}.\\
The dependence of these parameters on the isospin asymmetry (N/Z dependence ) displayed, in 
Fig.\ref{fig:7},  
is also
found to be in supportive nature with the findings in Fig.\ref{fig:6}.  
An increase in the number of neutrons will increase 
the number of collisions and hence dominance of R, $\frac{1}{Q_{zz}/nucleon}$, free particles as well as LCP's is
observed with increase in N/Z ratio. Nuclear stopping as well as LCP's are observed to be strongly dependent on the
isospin-dependent cross-section. Similar results with isospin dependent cross-section are observed in Figs.\ref{fig:3} 
and \ref{fig:4}. 
From here, one may conclude that the nuclear stopping and LCP's can also be used as a tool to 
investigate the isospin-dependent cross-section.\\
\begin{figure}
\includegraphics{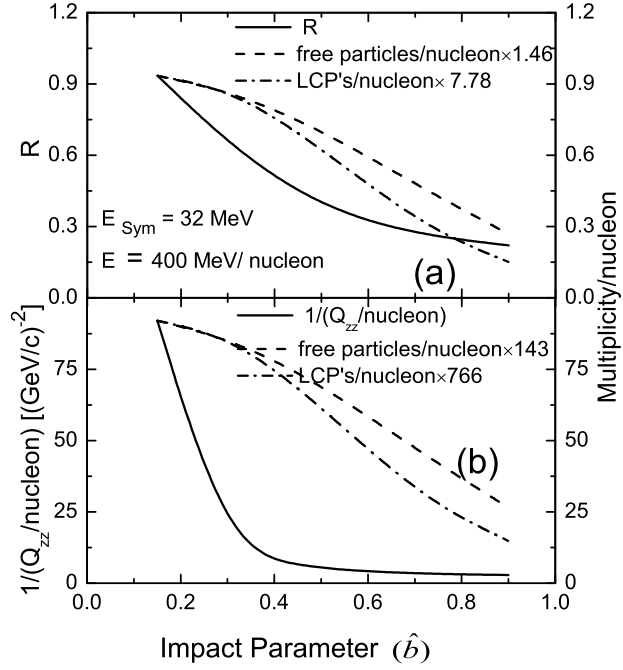}
\caption{\label{fig:8} The scaled free particles/nucleon and LCP's/nucleon along with
anisotropy ratio $R$ (a)
and $\frac{1}{Q_{zz}/nucleon}$(b) as a function of normalized impact parameter in the presence of symmetry energy. The reaction is at incident energy 400 MeV/nucleon.}
\end{figure}

To further elaborate this point, we display in Fig.\ref{fig:8},  
Multiplicity/nucleon (free and LCP's) as well as
$R$ and $\frac{1}{Q_{zz}/nucleon}$. Once free nucleons and LCP's are normalized
with $R$ at the starting point of impact parameter, we see that their behavior 
with respect to impact
parameter is similar to that of anisotropy ratio, whereas, visible difference occurs 
with reference to quadrupole moment. This similarity in all three quantities
in the presence of symmetry energy makes LCP's good indicator of global stopping in
heavy-ion collisions.\\

\section{Conclusion} 
In summary, using the isospin-dependent quantum molecular dynamics (IQMD)model, we investigate the emission of free particles, LCP's,
and degree of stopping reached in a heavy-ion collisions in the presence of symmetry energy
and isospin dependent cross-section. We observed that nuclear stopping
in term of anisotropy ratio and quadrupole moment depends weakly on the 
symmetry energy and strongly on the isospin-dependent cross-section. On the other hand, the symmetry energy and
isospin-dependent cross-section has an effect of $10\%$ on the LCP's production. It means nuclear stopping 
and LCP's production can be used as a tool to investigate the 
isospin-dependent cross-section. The LCP's production is found to be highly correlated with the global stopping.\\

\begin{acknowledgments}
This work has been supported by the Grant no. 03(1062)06/ EMR-II, from the Council of Scientific and
Industrial Research (CSIR) New Delhi, Govt. of India.\\
\end{acknowledgments}


\end{document}